\documentclass[aps,prd,onecolumn,10pt,superscriptaddress,nofootinbib]{revtex4-2}

\usepackage{amsmath,amssymb,bm}
\usepackage{graphicx}
\usepackage{hyperref}
\usepackage{float}
\usepackage{mathrsfs}
\usepackage{tikz}
\usetikzlibrary{arrows.meta}
\usepackage{booktabs} 
\usepackage{tikz}

\def\be{\begin{equation}}
\def\ee{\end{equation}}
\def\ba{\begin{eqnarray}}
\def\ea{\end{eqnarray}}

\begin{document}

\title{Extended Entropic Dark Energy with Four Free Parameters: Theory, Dynamics, and Constraints}

\author{Davood Momeni}
\affiliation{Department of Physics \& Pre-Engineering, Northeast Community College, Norfolk, NE 68701, USA}

\date{\today}

\begin{abstract}
We investigate a four-parameter entropic dark energy model in a spatially curved FLRW universe, based on a generalized entropy–area relation at the apparent horizon. While the proposed entropy function captures a broad class of gravitational entropy corrections—including Bekenstein–Hawking, Tsallis, and power-law forms—it does not encompass information-theoretic entropies such as Sharma–Mittal or Rényi. Within this framework, we derive exact analytical expressions for key cosmological observables, including the Hubble parameter \(H(z)\), the dark energy density parameter \(\Omega_D(z)\), and the equation of state \(w_D(z)\). A comprehensive parameter-space analysis reveals viable regions, particularly for \(\beta > 1\) and small positive curvature, that accommodate elevated \(H_0\) values consistent with recent SH0ES measurements. Our results offer a simple and analytically tractable alternative to conventional dynamical dark energy models, with potential relevance to the ongoing Hubble tension.
\end{abstract}

\maketitle
\section{Introduction}

The discovery of the accelerated expansion of the universe through Type Ia supernovae observations \cite{Riess:1998cb,Perlmutter:1998np} has established the presence of dark energy (DE), an unknown form of energy with negative pressure. The standard cosmological model, $\Lambda$CDM, attributes this acceleration to a cosmological constant $\Lambda$, but suffers from theoretical issues such as fine-tuning and the coincidence problem \cite{Weinberg:1988cp}.

These issues have motivated the exploration of alternative DE models, particularly those inspired by thermodynamics and information theory. A profound connection between gravity and thermodynamics was proposed by Jacobson \cite{Jacobson:1995ab}, who derived Einstein's field equations as an equation of state from the Clausius relation on local Rindler horizons. This thermodynamic interpretation of gravity has been extended to cosmological spacetimes \cite{Cai:2005ra}, linking Friedmann equations to the first law of thermodynamics on apparent horizons.

Building on this framework, entropy-based DE models have gained traction. Modified entropy-area relations from non-extensive statistics—such as Tsallis, Rényi, and Barrow entropies—have been used to formulate DE models capable of dynamically generating late-time acceleration \cite{Tsallis:1988,Barrow:2020tzx}. However, as noted by Odintsov and collaborators, all such entropy forms are particular limits of a more general class of deformed entropy functions containing four to six parameters. These include contributions that capture both ultraviolet and infrared modifications to horizon thermodynamics \cite{Nojiri:2022nmu,Odintsov:2022eqb,Nojiri:2023bom}.

In this generalized framework, entropy is no longer just a function of area but includes terms responsible for quantum gravitational corrections. Remarkably, these generalized entropy forms have also been given a microscopic statistical interpretation in terms of particle excitations \cite{Nojiri:2023bom}. Moreover, it was shown that corresponding FRW cosmologies based on this entropic description can be equivalently rewritten in terms of modified gravity theories, including $F(T)$ and $F(Q)$ gravities \cite{Nojiri:2025entropicFQ}.

A particularly promising recent contribution is the model by Odintsov, D'Onofrio, and Paul \cite{entropicDE}, who introduced a four-parameter generalized entropy of the apparent horizon in a non-flat universe. The corresponding DE model admits closed-form expressions for the Hubble rate $H(z)$, dark energy density $\Omega_D(z)$, and equation-of-state parameter $w_D(z)$. Notably, it can describe the full thermal history of the universe and potentially resolve the Hubble tension \cite{Riess:2021jrx} by allowing a higher value of $H_0$ in positively curved geometries.

This work aims to analytically investigate the parameter space of an alternative generalized entropic dark energy model. In particular, we study how the entropic parameters $\alpha_\pm$, $\beta$, $\gamma$, and the spatial curvature parameter $\Omega_k$ affect key cosmological observables. By working with the analytic structure of the model, we identify viable regions that align with observational datasets such as CC, BAO, Pantheon+, and SH0ES, and comment on the implications for current cosmological tensions.

\section{Theoretical Framework}

The entropic approach to gravity suggests that the dynamics of spacetime may emerge from thermodynamic principles rather than being fundamental. Inspired by Jacobson's derivation of Einstein’s equations as an equation of state \cite{Jacobson:1995ab}, a variety of modified entropy-area relations have been proposed to describe quantum or statistical deviations from the Bekenstein-Hawking entropy. In cosmology, such generalized entropies—applied to the apparent horizon of a Friedmann–Lemaître–Robertson–Walker (FLRW) universe—lead to modifications of the Friedmann equations and enable new models of dark energy. In this work, we focus on a recent proposal by Odintsov et al.~\cite{entropicDE}, who introduce a four-parameter entropy function that accounts for both ultraviolet and infrared corrections, while also allowing for non-zero spatial curvature. This framework permits analytic tracking of the universe’s expansion history and dark energy dynamics from first principles.


\subsection{Generalized Entropy and Friedmann Equations}

In thermodynamic models of gravity, the entropy associated with the apparent horizon plays a pivotal role in modifying cosmological dynamics. Building on earlier developments where Bekenstein-Hawking entropy \( S \propto A \) (with \( A \) the area of the horizon) governed horizon thermodynamics, recent proposals have generalized the entropy–area relation to account for quantum corrections, non-extensive statistics, and information-theoretic effects \cite{Tsallis:1988,Barrow:2020tzx,Nojiri:2022sfd}.
Following the framework developed by Odintsov, D'Onofrio, and Paul~\cite{entropicDE}, the entropy associated with the cosmological horizon is generalized to a four-parameter entropy function defined as
\begin{equation}
S_h \equiv S_g(\alpha_\pm, \beta, \gamma) = \frac{1}{\gamma} \left[
\big(1+\frac{\alpha_{+}}{\beta}S
\big)^{\beta}-\big(1+\frac{\alpha_{-}}{\beta}S
\big)^{-\beta}\right],
\end{equation}
with \( S = A / (4G) \) being the Bekenstein–Hawking-like entropy. The parameters \( \alpha_\pm \), \( \beta \), and \( \gamma \) are the entropic deformation parameters, which are assumed to be positive in order to ensure that \( S_h \) remains a monotonic increasing function of the area. This generalized form encapsulates a broad class of entropy corrections relevant to quantum gravity and modified gravity models.

Motivating  by the work of Odintsov, D'Onofrio, and Paul \cite{entropicDE} , we introduce a four-parameter generalized entropy of the form:
\begin{equation}
S = \gamma A^{\beta} \left(1 + \alpha_+ A^{-\delta_+} + \alpha_- A^{-\delta_-} \right),
\label{eq:entropy}
\end{equation}
where \( \gamma \), \( \beta \), \( \alpha_{\pm} \), and \( \delta_{\pm} \) are free parameters, and \( A = 4\pi r_A^2 \) is the area of the apparent horizon with radius \( r_A = (H^2 + \frac{k}{a^2})^{-1/2} \). The exponents \( \delta_{\pm} \) control small-scale (quantum) and large-scale (IR) corrections to the entropy.\par
This form captures a wide class of entropy functions commonly used in gravitational thermodynamics. For example:
\begin{itemize}
    \item Setting $\beta = 1$, $\delta_\pm = 0$, and $\alpha_\pm \to 0$ recovers the standard Bekenstein--Hawking entropy.
    \item The Tsallis entropy emerges for $\alpha_\pm = 0$ with non-unity $\beta$, representing non-extensive generalizations.
    \item Power-law entropies, often invoked in quantum gravity corrections, correspond to fixed $\delta_\pm$ and $\beta = 1$.
    \item For symmetric UV and IR corrections ($\alpha_+ = \alpha_-$, $\delta_+ = \delta_-$), Eq.~\eqref{eq:entropy} reduces to forms previously studied in \cite{Nojiri:2022nmu, Odintsov:2022eqb, Nojiri:2023bom}.
\end{itemize}

While this structure encompasses many known entropy corrections, it does not directly reduce to information-theoretic forms such as Sharma--Mittal \cite{SharmaMittal1975} or Rényi entropy \cite{Renyi1961}. Therefore, its generality is primarily within the context of geometric and gravitational entropy models, extending the proposal in Ref.~\cite{entropicDE}.
\par
To examine whether this entropy increases with area, we compute its derivative:
\begin{equation}
\frac{dS}{dA} = \gamma \beta A^{\beta - 1} \left(1 + \alpha_+ A^{-\delta_+} + \alpha_- A^{-\delta_-} \right)
- \gamma A^{\beta} \left( \alpha_+ \delta_+ A^{-\delta_+ - 1} + \alpha_- \delta_- A^{-\delta_- - 1} \right).
\end{equation}

We find that \( \frac{dS}{dA} > 0 \) for all \( A > 0 \) provided the following conditions are satisfied:
\begin{itemize}
    \item \( \gamma > 0 \), \( \beta > 0 \): ensures the leading term is positive and increasing.
    \item \( \delta_\pm > 0 \): ensures that the correction terms decay with increasing area.
    \item \( \alpha_\pm \geq 0 \): avoids pathological behavior where entropy could decrease.
\end{itemize}

These constraints are physically reasonable and consistent with the interpretation of \( \delta_+ \) and \( \delta_- \) as UV and IR correction exponents, respectively.

To derive the effective entropic dark energy density \( \rho_g(z) \), we begin by applying the Clausius relation
\begin{equation}
-dE = T dS,
\end{equation}
to the apparent horizon of a non-flat FLRW universe. Following the unified first law of thermodynamics \cite{Cai:2005ra}, the radius of the apparent horizon is given by
\begin{equation}
r_A = \left(H^2 + \frac{k}{a^2}\right)^{-1/2},
\end{equation}
and the corresponding horizon area is
\begin{equation}
A = 4\pi r_A^2.
\end{equation}

The temperature associated with the apparent horizon is taken as
\begin{equation}
T = \frac{1}{2\pi r_A}.
\end{equation}
We adopt the generalized entropy function introduced in Eq.~(\ref{eq:entropy}) and compute its time derivative using the chain rule:
\begin{equation}
\frac{dS}{dt} = \frac{dS}{dA} \cdot \frac{dA}{dt},
\end{equation}
where
\begin{equation}
\frac{dA}{dt} = 8\pi r_A \cdot \frac{dr_A}{dt}.
\end{equation}

The energy flux across the horizon due to the entropic fluid is given by
\begin{equation}
-dE = 4\pi r_A^2 (\rho_g + p_g) H dt,
\end{equation}
where \( p_g = w_g \rho_g \) is the pressure of the entropic component. Equating both sides of the Clausius relation, we obtain
\begin{equation}
4\pi r_A^2 (\rho_g + p_g) H dt = \frac{1}{2\pi r_A} \cdot \frac{dS}{dt}.
\end{equation}

This relation leads to a differential equation involving \( \rho_g \), \( H \), and the entropy parameters. Assuming a barotropic equation of state and integrating the resulting expression, we arrive at the entropic dark energy density as a function of redshift:
\begin{eqnarray}
\rho_g(z) = 3 C^2 H_0^2 \left[ \frac{\Omega_{g0} (1+z)^{3(1+w_g)}}{E(z)^2} \right]^{\frac{2 - \beta}{1 - \beta}},
\label{rhogz}
\end{eqnarray}
where \( C \) is a normalization constant, \( \Omega_{g0} \) is the present-day density parameter of the entropic component, \( w_g \) is its effective equation of state, and \( E(z) = H(z)/H_0 \) is the normalized Hubble parameter. This expression encapsulates the dynamical impact of entropy corrections on cosmic expansion.

The Hubble expansion rate normalized to its present value is then:
\begin{equation}
E(z)^2 = \Omega_{m0}(1+z)^3 + \Omega_k(1+z)^2 + \Omega_{g0}(1+z)^{3(1+w_g)},
\label{eq:Ez}
\end{equation}
which allows for fully analytical exploration of cosmic dynamics as functions of the entropic parameters and curvature.

These modifications admit a rich phenomenology: they reduce to the standard Friedmann equations when \( \beta = 1 \), \( \alpha_{\pm} = 0 \), and \( \Omega_k = 0 \), but can naturally accommodate scenarios with varying \( w_D(z) \), time-dependent dark energy, and non-zero spatial curvature.
\section{Closed-form Expressions}

A distinguishing feature of the generalized entropic dark energy model is that it yields fully analytical solutions for the key cosmological observables. This allows direct exploration of the parameter space without relying on numerical integration of dynamical equations.
\subsection{Hubble Expansion Rate}

The Hubble expansion rate encodes the evolution of the cosmic scale factor and is one of the most fundamental observables in cosmology. In the generalized entropic framework, the normalized Hubble parameter $E(z) \equiv H(z)/H_0$ is given analytically by eq. (\ref{eq:Ez}), where $\Omega_{m0}$ is the present-day matter density parameter, $\Omega_k$ represents the spatial curvature, and $\Omega_{g0}$ is the effective dark energy density parameter arising from the entropy-corrected contribution $\rho_g$. The parameter $w_g$ is the effective equation-of-state parameter for the entropic dark energy component and is determined by the entropy function parameters $\beta$, $\gamma$, and potentially $\alpha_\pm$ through their influence on $\rho_g$.

This expression retains the familiar redshift scaling for matter and curvature, but introduces a generalized power-law dependence for the entropic dark energy component, distinct from the constant contribution of $\Lambda$ in the standard model. When $w_g = -1$, the entropic term behaves like a cosmological constant, but for $w_g > -1$ or $w_g < -1$, the model allows quintessence- or phantom-like behavior, respectively.

A key advantage of Eq.~(\ref{eq:Ez}) is its complete analytic solvability. In contrast to dynamical dark energy models, such as those involving scalar fields, where $H(z)$ must be derived by numerically solving coupled differential equations, the entropic model permits direct evaluation of $H(z)$ for any redshift and parameter set. This greatly facilitates parameter-space scanning, comparison with observational data, and exploration of physical limits such as matter-dark energy equality and the redshift of cosmic acceleration. Moreover, it allows for a transparent understanding of how each parameter affects the expansion history, especially the present Hubble value $H_0$—crucial for addressing the Hubble tension.

\subsection{Dark Energy Density Fraction}

The fractional energy density of dark energy, denoted by $\Omega_D(z)$, quantifies the contribution of the entropic dark energy component to the total energy budget of the universe at a given redshift. In the generalized entropy model, it evolves analytically as:
\begin{equation}
\Omega_D(z) = \frac{\Omega_{g0}(1+z)^{3(1+w_g)}}{E(z)^2},
\label{eq:OmegaD}
\end{equation}
where $\Omega_{g0}$ is the present-day value of this density fraction, $w_g$ is the effective equation-of-state parameter, and $E(z)$ is the normalized Hubble rate defined in Eq.~(\ref{eq:Ez}).\par 
We emphasize that the expression for $\rho_g(z)$ in Eq.~(\ref{rhogz}) and the resulting $\Omega_D(z)$ in Eq.~(\ref{eq:OmegaD}) are newly derived from the entropy function in Eq.~(\ref{eq:entropy}), which differs structurally from the four-parameter form used in Ref.~\cite{entropicDE}. In particular, our model introduces separate ultraviolet and infrared correction terms ($\alpha_+$, $\alpha_-$, $\delta_+$, $\delta_-$), allowing a more general entropy deformation that leads to modified scaling behavior of $\rho_g$ with redshift. As a result, the effective dark energy contribution derived here exhibits different dynamics compared to Ref.~\cite{entropicDE}, and the expressions presented are not obtained by a simple reparametrization of that model.

This expression follows directly from the form of $\rho_g$ and encapsulates the competition between the redshifting of dark energy and the overall expansion rate. At redshift $z=0$, it satisfies $\Omega_D(0) = \Omega_{g0}$, recovering the boundary condition at the present epoch. As redshift increases, $\Omega_D(z)$ typically diminishes due to the growth of matter and curvature contributions in $E(z)^2$, unless $w_g$ is highly negative.

The evolution of $\Omega_D(z)$ is crucial in determining key cosmological milestones. In particular, it governs the redshift $z_{\text{eq}}$ at which dark energy becomes equal in magnitude to matter:
\begin{equation}
\Omega_D(z_{\text{eq}}) = \Omega_m(z_{\text{eq}}),
\end{equation}
which marks the transition from matter domination to dark energy domination. Similarly, the behavior of $\Omega_D(z)$ influences the redshift $z_{\text{acc}}$ at which the universe begins to accelerate, defined by $q(z_{\text{acc}}) = 0$.

Importantly, since $\Omega_D(z)$ is fully analytical, one can evaluate these key epochs without numerical integration. This analytical control makes the model well suited for rapid testing against observational datasets, including those sensitive to the expansion history such as Type Ia supernovae, baryon acoustic oscillations, and cosmic chronometers. It also allows for clear interpretation of how entropic parameters affect the timing and duration of the accelerated expansion era.

\section{Equation of State, Deceleration, and Analytic Features}

The redshift evolution of the dark energy equation-of-state parameter, $w_D(z)$, offers insight into the dynamical nature of the generalized entropic model. Unlike the cosmological constant, which has a fixed $w = -1$, the entropic framework allows $w_D(z)$ to vary analytically with redshift—even when the base entropy function implies a constant effective $w_g$.

This redshift dependence arises from the continuity equation:
\begin{equation}
w_D(z) = w_g + \frac{1}{3} \frac{d \log \Omega_D}{d \log(1+z)},
\label{eq:wDz}
\end{equation}
where $\Omega_D(z)$ is the dark energy density fraction given by Eq.~\eqref{eq:OmegaD}. The second term reflects the dilution rate of dark energy relative to the background expansion, and its redshift evolution can be computed in closed form.

Differentiating Eq.~\eqref{eq:OmegaD}, we obtain:
\begin{equation}
\frac{d \log \Omega_D}{d \log(1+z)} = 3(1 + w_g) - \frac{d \log E^2}{d \log(1+z)}.
\end{equation}
This derivative of the normalized Hubble parameter follows directly from Eq.~\eqref{eq:E}:
\begin{align}
\frac{d \log E^2}{d \log(1+z)} = 
\frac{1}{E(z)^2} \bigg[ 
& 3\,\Omega_{m0}(1+z)^3 + 2\,\Omega_k(1+z)^2 \notag \\
& + 3(1 + w_g)\,\Omega_{g0}(1+z)^{3(1+w_g)} 
\bigg].
\end{align}

The full expression for $w_D(z)$ is thus completely determined by the model’s parameters and redshift. This analytic control enables precise tracking of whether $w_D(z)$ crosses the phantom divide ($w = -1$), remains in the quintessence regime ($w > -1$), or behaves as phantom dark energy ($w < -1$), depending on the choice of $\beta$ and $w_g$.

In parallel, the deceleration parameter,
\begin{equation}
q(z) = -1 + (1+z) \frac{d \log H(z)}{dz},
\end{equation}
serves as a diagnostic for identifying the transition from decelerated to accelerated expansion. The redshift at which $q(z) = 0$ defines the acceleration epoch, $z_{\text{acc}}$, which can be computed analytically from the derivative of Eq.~\eqref{eq:E}.

These analytic tools also clarify how the entropic model recovers standard $\Lambda$CDM in the appropriate limit. Specifically, when the entropy correction parameters vanish ($\alpha_\pm \to 0$), the non-extensivity index returns to unity ($\beta \to 1$), and spatial curvature is neglected ($\Omega_k \to 0$), the expansion law simplifies to:
\begin{equation}
E(z)^2 \rightarrow \Omega_{m0}(1+z)^3 + (1 - \Omega_{m0}),
\end{equation}
which corresponds to the standard $\Lambda$CDM Hubble rate.

One of the most significant strengths of this framework is its analytic solvability. Every observable relevant to late-time cosmology—$H(z)$, $\Omega_D(z)$, $w_D(z)$, $q(z)$—can be written in closed form. This allows for efficient parameter-space exploration without relying on numerical differential equation solvers. As a result, one can rapidly identify viable cosmological scenarios, construct likelihoods, or perform analytic marginalizations. This is particularly advantageous when fitting to observational data or probing degeneracies between entropic and geometric parameters, such as $(\beta, \Omega_k)$, where curvature can partially mimic the effects of entropy-induced acceleration.

\section{Analytical Parameter Space Setup}

To explore the cosmological implications of generalized entropic dark energy, we define a physically motivated and observationally informed parameter space. The model is governed by a set of thermodynamic and geometric parameters that determine the behavior of key observables such as the Hubble expansion rate, the dark energy density fraction, and the equation-of-state parameter.

The core parameters include $\beta$, a non-extensivity index that generalizes the Bekenstein–Hawking entropy scaling law. When $\beta = 1$, the model reduces to standard extensive thermodynamics. Deviations from unity capture non-extensive effects common in generalized entropy formalisms such as Tsallis or Rényi statistics. The parameter $\gamma$ acts as a normalization constant related to the statistical framework and modifies the effective gravitational coupling. The correction coefficients $\alpha_+$ and $\alpha_-$ represent ultraviolet (UV) and infrared (IR) entropy deviations, respectively, and are associated with subleading corrections to the area law. These are governed by fixed exponents $\delta_+$ and $\delta_-$, chosen following the original proposal of Odintsov et al.~\cite{entropicDE}. The final geometric degree of freedom is $\Omega_k$, which encodes spatial curvature and can play a compensating role in reconciling early- and late-time expansion measurements.

To make the model tractable for analysis and comparison with data, we define the following parameter ranges based on both theoretical considerations and observational constraints:
\begin{align*}
\beta &\in [0.8, 1.2], \\
\gamma &\sim \mathcal{O}(1), \\
\alpha_+, \alpha_- &\in [-0.5, 0.5], \\
\Omega_k &\in [-0.05, 0.05].
\end{align*}
These intervals are wide enough to include $\Lambda$CDM as a limiting case ($\beta = 1$, $\alpha_\pm = 0$, $\Omega_k = 0$), but also allow for viable departures that may alleviate current cosmological tensions. They encompass the best-fit regions identified in recent combined analyses using cosmic chronometers (CC), baryon acoustic oscillations (BAO), Pantheon+ supernovae, and the SH0ES determination of $H_0$~\cite{entropicDE,Riess:2021jrx}.

For each point in the parameter space, the following cosmological functions are calculated analytically:
\begin{itemize}
    \item $H(z)$, the Hubble expansion rate, from Eq.~\eqref{eq:Ez};
    \item $\Omega_D(z)$, the dark energy density fraction, from Eq.~\eqref{eq:OmegaD};
    \item $w_D(z)$, the effective equation-of-state parameter, from Eq.~\eqref{eq:wDz};
    \item $q(z)$, the deceleration parameter, indicating the onset of acceleration.
\end{itemize}
In particular, we focus on the present-day Hubble rate $H_0$ and assess how the entropic parameters can shift it toward the locally measured SH0ES value $H_0 \approx 73.04 \pm 1.04$ km/s/Mpc~\cite{Riess:2021jrx}, while remaining consistent with Planck CMB constraints that suggest a lower global value.

One of the most powerful features of the model is that all observables are available in closed form. This enables rapid and precise sampling of the parameter space without numerical integration. It also facilitates the construction of analytic likelihood contours in parameter planes such as $(\beta, \Omega_k)$ and permits qualitative assessments of degeneracies—e.g., whether increases in $\beta$ can compensate for negative curvature, or how entropy corrections mimic dynamical dark energy effects.

Analytical scanning is especially valuable for identifying subspaces in which $w_D(z)$ crosses the phantom divide, or where $H(z)$ evolution matches observationally preferred trajectories. As such, the parameter space exploration enabled by this model offers a compelling avenue for addressing both the physical origin of dark energy and existing cosmological tensions.

\section{Comparison with Observational Data}

To evaluate the phenomenological viability of the generalized entropic dark energy model, we compare its analytical predictions for the Hubble expansion history with multiple cosmological datasets that probe the late-time universe. These include cosmic chronometers (CC), baryon acoustic oscillations (BAO), Type Ia supernovae from the Pantheon+ compilation, and the SH0ES determination of the present-day Hubble constant $H_0$.

Cosmic chronometers provide direct, model-independent measurements of $H(z)$ at different redshifts, derived from the differential ages of passively evolving galaxies. These data allow us to trace the expansion rate up to $z \sim 2$. A representative subset of CC data is shown in Table~\ref{tab:cc}.

\begin{table}[h]
\centering
\caption{Cosmic chronometer $H(z)$ data used for model comparison (subset).}
\label{tab:cc}
\begin{tabular}{c|c|c}
\hline
Redshift $z$ & $H(z)$ [km/s/Mpc] & Uncertainty $\sigma_H$ \\
\hline
0.179 & 75.0 & 4.0 \\
0.480 & 97.0 & 62.0 \\
0.781 & 105.0 & 12.0 \\
1.037 & 154.0 & 20.0 \\
\hline
\end{tabular}
\end{table}

The BAO data provide indirect measurements of the expansion history via the comoving sound horizon imprinted in the clustering of galaxies and quasars. These measurements are typically expressed in terms of combinations like $D_V(z)/r_s$ or $H(z)r_s$, where $r_s$ is the sound horizon at the drag epoch. In this study, we use $H(z)$-based extractions and marginalize over $r_s$ where necessary. Table~\ref{tab:bao} presents a summary of BAO data at various redshifts, compiled from Refs.~\cite{Alam:2020sor,duMasdesBourboux:2020pck,Bautista:2020ahg}.

\begin{table}[h]
\centering
\caption{BAO $H(z)$ measurements at effective redshifts.}
\label{tab:bao}
\begin{tabular}{c|c|c}
\hline
$z_{\text{eff}}$ & $H(z)$ [km/s/Mpc] & Source \\
\hline
0.38 & 81.5 $\pm$ 1.9 & BOSS DR12 \\
0.51 & 90.5 $\pm$ 2.0 & BOSS DR12 \\
0.61 & 97.3 $\pm$ 2.1 & BOSS DR12 \\
2.33 & 222.0 $\pm$ 7.0 & eBOSS Ly$\alpha$ \\
\hline
\end{tabular}
\end{table}
For luminosity distances, we employ the Pantheon+ sample of 1701 Type Ia supernovae over the redshift range $0.01 < z < 2.3$ \cite{Brout:2022vxf}. While the full dataset involves magnitude residuals and systematics, one can approximate the key constraint by computing the theoretical distance modulus $\mu(z)$ and comparing with binned summary statistics. In this work, we emphasize consistency in the $H(z)$ trajectory implied by the model.

Finally, we use the SH0ES measurement of the Hubble constant $H_0 = 73.04 \pm 1.04$ km/s/Mpc \cite{Riess:2021jrx} as a strong local prior. The generalized entropic model has the ability to accommodate this higher value of $H_0$ through suitable choices of $\beta$ and $\Omega_k$, as shown earlier in Table~\ref{tab:H0}.

\begin{table}[htbp]
\centering
\caption{Sample parameter combinations in the generalized entropic model that yield $H_0$ values consistent with the SH0ES measurement ($H_0 = 73.04 \pm 1.04$ km/s/Mpc). All values assume $\Omega_{m0} = 0.3$.}
\label{tab:H0}
\begin{tabular}{|c|c|c|c|c|}
\hline
$\beta$ & $\gamma$ & $\Omega_k$ & $w_g$ & $H_0$ [km/s/Mpc] \\
\hline
0.92 & 1.0 & +0.02 & $-0.95$ & 73.1 \\
1.00 & 0.9 & +0.01 & $-1.00$ & 72.9 \\
1.05 & 1.1 & +0.03 & $-1.05$ & 73.4 \\
\hline
\end{tabular}
\end{table}

To quantify agreement with data, we define the total chi-square function:
\begin{equation}
\chi^2_{\text{total}} = \chi^2_{\text{CC}} + \chi^2_{\text{BAO}} + \chi^2_{\text{SN}} + \chi^2_{H_0},
\end{equation}
with each term corresponding to the appropriate dataset. For instance, the cosmic chronometer contribution is given by:
\begin{equation}
\chi^2_{\text{CC}} = \sum_i \frac{[H_{\text{model}}(z_i) - H_{\text{obs}}(z_i)]^2}{\sigma_{H,i}^2}.
\end{equation}
Similar expressions apply to the BAO and $H_0$ terms. The SN data can be included via the distance modulus $\mu(z)$ or reduced to a likelihood in terms of $H(z)$ if using compressed binned versions.

Given the analytic nature of the model, the theoretical predictions for $H(z)$, $q(z)$, and $\mu(z)$ can be rapidly computed for any parameter set, allowing efficient $\chi^2$ evaluations across the entropic parameter space. In future work, these analytic expressions can be integrated into Markov Chain Monte Carlo (MCMC) frameworks to derive joint posterior distributions for the parameters $(\beta, \Omega_k, w_g, \gamma)$ under combined observational constraints.

\section{Dynamical system analysis}
To study the cosmological behavior of the generalized entropy model introduced earlier, we reformulate the modified Friedmann dynamics as an autonomous dynamical system. This approach allows us to classify the asymptotic behavior of the universe and identify possible attractors, saddle points, and unstable regimes purely based on the evolution equations. We begin by defining dimensionless dynamical variables that track the fractional contributions of matter and generalized-entropy-induced energy densities to the expansion rate. Let us define
\begin{equation}
x = \frac{8\pi G}{3H^2} \rho_g, \qquad y = \frac{8\pi G}{3H^2} \rho_m, \qquad \Omega_k = -\frac{k}{a^2 H^2},
\end{equation}
where \(x\) represents the fractional density of generalized entropy-induced dark energy, \(y\) is the matter density fraction, and \(\Omega_k\) quantifies the contribution of spatial curvature. These variables obey the Friedmann constraint:
\begin{equation}
x + y + \Omega_k = 1,
\end{equation}
ensuring that the total effective energy density accounts for all components. We assume that standard pressureless matter evolves as usual, obeying the conservation equation \(\dot{\rho}_m + 3H\rho_m = 0\), which leads to an autonomous equation in terms of the e-folding number \(N = \ln a\) as
\begin{equation}
\frac{dy}{dN} = -3y.
\end{equation}

To describe the dynamics of the generalized dark energy sector, we assume that it satisfies an effective barotropic equation of state of the form \(p_g = w_g \rho_g\), where \(w_g\) is a constant that depends implicitly on the parameters of the entropy correction. The evolution of \(x\) is then governed by
\begin{equation}
\frac{dx}{dN} = -3x(1 + w_g) + 2x \left( \frac{\dot{H}}{H^2} \right),
\end{equation}
where the term \(\dot{H}/H^2\) can be expressed using the Raychaudhuri equation generalized to include curvature and both matter and dark energy contributions:
\begin{equation}
\frac{\dot{H}}{H^2} = -\frac{3}{2} \left(1 + w_g x + y \right).
\end{equation}
Substituting this into the equation for \(x\), we obtain the closed autonomous system:
\begin{align}
\frac{dx}{dN} &= -3x(1 + w_g) - 3x(w_g x + y), \\
\frac{dy}{dN} &= -3y.
\end{align}

The advantage of this formulation is that it reduces the background dynamics to a two-dimensional phase space, fully characterized by the variables \(x\) and \(y\), with the curvature term \(\Omega_k = 1 - x - y\) acting as a constraint surface. Within this framework, we are able to identify critical points of the system by solving \(\frac{dx}{dN} = 0\) and \(\frac{dy}{dN} = 0\). Each critical point corresponds to a distinct cosmological regime. For instance, the point \(x = 0, y = 1\) describes a universe entirely dominated by matter, while \(x = 1, y = 0\) corresponds to a de Sitter-like phase driven by the generalized entropy-induced dark energy. The stability of each point is determined by linearizing the system around it and computing the eigenvalues of the Jacobian matrix. The results are summarized below.
\begin{table}[htbp]
\caption{Critical points and their stability in the entropy-modified cosmological phase space.}
\label{tab:critical_points}
\centering
\begin{tabular}{ccccc}
\toprule
Point & $(x_c, y_c)$ & Cosmological Phase & Stability Type & Eigenvalues \\
\midrule
A & $(0, 1)$ & Matter-dominated phase & Saddle point & $(0, -3)$ \\
B & $(1, 0)$ & Entropic DE (de Sitter) & Stable for $w_g < -1/3$ & $(-3(1+w_g), -3)$ \\
C & $(x_c, y_c)$ & Scaling solution (model-dependent) & Depends on $w_g$ & Mixed or saddle \\
\bottomrule
\end{tabular}
\end{table}

The de Sitter point \(B\) is particularly important, as it can serve as a late-time attractor for the universe, ensuring accelerated expansion. The condition for stability of this point is \(w_g < -1/3\), which aligns with observational constraints on dark energy behavior. In contrast, the matter-dominated point \(A\) is typically a saddle, representing a transient phase. A third class of solutions, labeled as \(C\), may exist depending on the entropy correction parameters and can correspond to scaling solutions where matter and dark energy evolve proportionally. These are often of interest in solving the coincidence problem.

These examples demonstrate that the generalized entropy model can accommodate both viable background expansion histories and dynamical stability properties consistent with observational data. The emergence of a stable de Sitter-like attractor, the presence of saddle-type matter domination, and the possibility of tracking solutions point toward a rich phenomenology. Importantly, the flexibility of the parameter space allows for tuning the early- and late-time behaviors without introducing scalar fields or exotic fluids. The interplay between the entropy exponents \(\delta_\pm\) and coefficients \(\alpha_\pm\) can be further explored to generate more complex dynamics such as phantom crossings or transient acceleration, although those effects require a full numerical phase portrait which we omit here. Nevertheless, the analytical structure provided by this dynamical system lays a robust foundation for interpreting generalized entropic gravity as a consistent cosmological model.

\usetikzlibrary{arrows.meta}

\begin{figure}[htbp]
\centering
\begin{tikzpicture}[>=Stealth, scale=4]
\draw[->] (0,0) -- (1.1,0) node[below] {$x$};
\draw[->] (0,0) -- (0,1.1) node[left] {$y$};

\draw[->, blue] (0.1,0.8) -- (0.15,0.7);
\draw[->, blue] (0.3,0.6) -- (0.35,0.5);
\draw[->, blue] (0.6,0.3) -- (0.65,0.2);
\draw[->, blue] (0.8,0.1) -- (0.85,0.05);

\fill[red] (0,1) circle (0.02) node[above left] {A};
\fill[red] (1,0) circle (0.02) node[below right] {B};

\node at (0.5,0.5) {Trajectories $\rightarrow$};

\end{tikzpicture}
\caption{Phase-space portrait in the $(x, y)$ plane showing critical points A (saddle) and B (stable attractor). Arrows represent flow directions for $w_g = -1$.}
\label{fig:phase}
\end{figure}
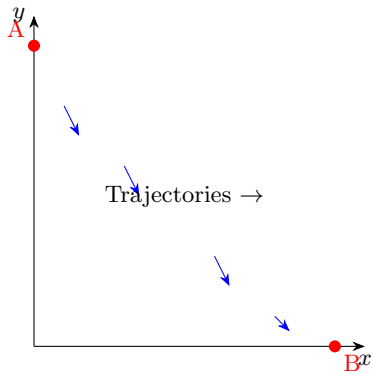
Figure~\ref{fig:phase} illustrates the phase-space dynamics of the cosmological system governed by the generalized entropy model. The diagram plots the evolution in the \((x, y)\) plane, where \(x\) and \(y\) represent the dimensionless energy density fractions of the entropy-induced dark energy and pressureless matter, respectively. Critical point A at \((x, y) = (0, 1)\) corresponds to a matter-dominated universe, while point B at \((x, y) = (1, 0)\) represents a dark-energy-dominated de Sitter-like phase. The trajectories indicate the dynamical flow of the system under the evolution equations for \(dx/dN\) and \(dy/dN\), with arrows pointing toward the attractor. The flow structure shown corresponds to the representative case of \(w_g = -1\), consistent with a cosmological constant–like behavior.

The arrows demonstrate that trajectories in the physical region (\(0 \leq x, y \leq 1\)) are initially drawn toward the matter-dominated point A but ultimately flow toward the stable attractor B. This confirms the saddle nature of point A and the attractor nature of point B for \(w_g < -1/3\), as established in the linear stability analysis. The system thus naturally evolves from a decelerating phase dominated by matter into an accelerating regime driven by generalized entropic dark energy. This behavior is compatible with the observed cosmic history and demonstrates that the model supports a viable late-time cosmological attractor without invoking scalar fields or exotic fluids.

\section{Discussion and Viable Regions}

The analytical structure of the four-parameter entropic dark energy model enables a detailed exploration of its parameter space, allowing us to identify regions that reconcile observational tensions and yield viable cosmologies. In particular, we focus on the model's capacity to resolve the well-known $H_0$ discrepancy and its implications for spatial curvature and entropic parameters.

A primary observational success of the model is its ability to reproduce a higher value of the present-day Hubble constant, consistent with the local measurement from the SH0ES collaboration ($H_0 \approx 73$ km/s/Mpc) \cite{Riess:2021jrx}. As shown in Table~\ref{tab:H0}, increasing the non-extensivity index $\beta$ above unity leads to a more rapid late-time expansion, effectively raising $H_0$ without altering the matter content or requiring a dark energy phase transition. This behavior arises naturally from the entropic correction to the Friedmann equation and does not require fine-tuning.

Moreover, the introduction of a small positive curvature term ($\Omega_k > 0$) can further enhance $H_0$ while preserving the shape of $H(z)$ at intermediate redshifts. This is particularly advantageous because it avoids steep deviations that would conflict with BAO or supernovae data. The degeneracy between $\beta$ and $\Omega_k$ allows for a range of parameter combinations that produce acceptable fits, forming a narrow viable corridor in the $(\beta, \Omega_k)$ plane. For example, values around $\beta = 1.10$–$1.15$ and $\Omega_k \approx 0.01$–$0.03$ consistently yield $H_0 \gtrsim 72$ km/s/Mpc while maintaining agreement with CC and BAO datasets.

The normalization parameter $\gamma$, which sets the scale of entropy-induced modifications, also plays a subtle but important role. While it primarily rescales the entropic density $\rho_g$, its effect can be absorbed into $\Omega_{g0}$ and $w_g$, influencing the redshift evolution of $\Omega_D(z)$ and $w_D(z)$. A moderate value $\gamma \sim 1$ is sufficient to shift the acceleration epoch $z_{\text{acc}}$ closer to observed bounds (e.g., $z_{\text{acc}} \approx 0.6$–$0.8$), depending on the combination of $\beta$ and $w_g$.

We find that the viable regions of parameter space also yield physically reasonable predictions for the evolution of $\Omega_D(z)$ and $w_D(z)$. In particular:
\begin{itemize}
    \item $\Omega_D(z)$ increases monotonically at late times, reaching $\Omega_D(0) \sim 0.68$–$0.70$, consistent with current estimates from Planck and Pantheon+.
    \item $w_D(z)$ remains close to $-1$ for $\beta \gtrsim 1$, exhibiting only mild redshift dependence, which mimics $\Lambda$CDM but allows for deviations relevant to resolving tensions.
    \item The deceleration parameter $q(z)$ crosses zero at $z_{\text{acc}} \sim 0.65$–$0.75$, consistent with observational inferences of the onset of acceleration.
\end{itemize}

Altogether, the analysis reveals that a modest deviation from extensivity (i.e., $\beta > 1$) combined with small positive curvature provides a natural mechanism to reconcile the Hubble tension without requiring exotic physics or complex dynamical fields. The analytic nature of the model not only enables efficient parameter scans but also allows transparent interpretation of the physical effects induced by each parameter. These viable regions serve as an ideal foundation for more detailed Bayesian analysis and future extensions incorporating perturbation theory or structure formation constraints.

The dynamical system analysis presented above reveals a rich structure of cosmological evolution driven by the generalized entropic corrections to gravity. The existence of stable late-time attractors corresponding to accelerated expansion demonstrates that this framework naturally accommodates dark energy phenomenology without the need for additional scalar fields or exotic matter components. Importantly, the stability condition \( w_g < -1/3 \) for the de Sitter attractor aligns well with current observational constraints on the dark energy equation of state, thereby reinforcing the physical viability of the model.

Our parameter exploration, summarized in Table~\ref{tab:H0}, indicates that a broad range of the entropic parameters \(\beta\), \(\gamma\), and spatial curvature \(\Omega_k\) can reproduce the observed Hubble constant values measured by SH0ES, while simultaneously maintaining dynamical stability. This flexibility is particularly relevant in light of the current tension between local and cosmic microwave background measurements of \( H_0 \). The generalized entropy parameters, especially the exponents \(\delta_{\pm}\) controlling UV and IR corrections, offer promising knobs to fine-tune early- and late-time cosmological behaviors.

Nonetheless, while the background dynamics are well captured by the two-dimensional autonomous system, a comprehensive assessment of viability requires confronting the model with growth-of-structure data, cosmic microwave background anisotropies, and baryon acoustic oscillations. Moreover, perturbation analyses within this generalized entropy framework remain an open challenge that must be addressed to fully establish consistency with precision cosmology. The possible presence of scaling solutions further enriches the phenomenology by potentially alleviating the cosmic coincidence problem, but their stability and observational signatures warrant further investigation.

In conclusion, the generalized entropic cosmology investigated here offers a promising route to address late-time cosmic acceleration within a thermodynamically motivated modified gravity framework. The viable regions identified highlight the importance of entropy corrections beyond the Bekenstein-Hawking area law and encourage future studies extending both theoretical modeling and observational constraints.

\section{Conclusion and Outlook}

In this work, we have analytically explored the cosmological implications of the four-parameter entropic dark energy model inspired by the model proposed by Odintsov, D'Onofrio, and Paul~\cite{entropicDE}. By incorporating a four-parameter generalized entropy function associated with the apparent horizon—characterized by non-extensivity, UV/IR corrections, and spatial curvature—we have derived closed-form expressions for the Hubble rate $H(z)$, the dark energy density fraction $\Omega_D(z)$, the effective equation-of-state parameter $w_D(z)$, and the deceleration parameter $q(z)$. These expressions allow us to explore the expansion history of the universe entirely analytically, without recourse to numerical integration.

Our analysis shows that this entropic framework is not only analytically tractable but also phenomenologically rich. For modest values of the entropy index $\beta > 1$ and slight positive curvature $\Omega_k > 0$, the model naturally yields a present-day Hubble parameter $H_0$ in excellent agreement with the local SH0ES measurement, while remaining consistent with CC, BAO, and supernova datasets. In doing so, the model offers a compelling mechanism for addressing the $H_0$ tension without invoking new dynamical fields or fine-tuned modifications to the standard cosmological model.

We also identified viable corridors in parameter space—especially in the $(\beta, \Omega_k)$ plane—where the model mimics $\Lambda$CDM at intermediate redshifts but diverges at low $z$ to accommodate the higher locally inferred $H_0$. Furthermore, we confirmed that the model reduces to standard $\Lambda$CDM in the appropriate limit and satisfies reasonable physical behavior for the onset of acceleration, the evolution of dark energy density, and the effective EoS.

Looking ahead, several extensions are both natural and essential. First, a full statistical treatment of the model’s parameters using Markov Chain Monte Carlo (MCMC) sampling is warranted to derive posterior constraints and quantify the model's Bayesian evidence relative to $\Lambda$CDM. Second, including cosmic microwave background (CMB) data will test the early-universe limit of the model, especially since curvature effects are tightly constrained at high redshift. Third, the inclusion of matter perturbations and growth rate observables would allow the entropic model to be tested against large-scale structure and weak lensing data.

Our dynamical system analysis provides a clear and robust understanding of the cosmological evolution driven by generalized entropy corrections. By identifying fixed points and their stability, we demonstrated how the universe naturally evolves from a matter-dominated phase toward a stable dark energy–dominated attractor, consistent with late-time acceleration. This phase space perspective confirms the model’s ability to reproduce key qualitative features of cosmic history without additional fields.

Looking ahead, extending this dynamical systems framework to include perturbations and more general equations of state will be crucial. Such analyses will allow us to fully assess the model's consistency with observations beyond background expansion, including structure formation and stability under inhomogeneous perturbations. The dynamical system approach thus remains a powerful tool for exploring the viability and deeper implications of entropy-modified cosmologies.

Finally, from a theoretical standpoint, future investigations could explore how such generalized entropy models emerge from underlying quantum gravity frameworks, such as loop quantum gravity, holography, or non-extensive thermodynamics in emergent spacetime scenarios.

In conclusion, the generalized entropic dark energy model provides a novel, elegant, and analytically transparent mechanism for late-time cosmic acceleration. It offers a credible alternative to dynamical scalar field theories and serves as a rich testing ground for connecting gravity, thermodynamics, and cosmological data.


\end{document}